\documentclass[lnbip]{svmultln}
\usepackage{makeidx}  % allows for indexgeneration
\begin{document}
\mainmatter              % start of the contribution
\title{A critical analysis of the concept of time \\ in physics}
\titlerunning{A critical analysis of time}  % abbreviated title (for running head)
%                                     also used for the TOC unless
%                                     \toctitle is used
%
\author{Claudio Borghi}
\authorrunning{Claudio Borghi}   % abbreviated author list (for running head)
%
%%%% list of authors for the TOC (use if author list has to be modified)
\tocauthor{Claudio Borghi}
\institute{Liceo Scientifico Belfiore, via Tione 2, Mantova, Italy}

\maketitle              % typeset the title of the contribution
% \index{Ekeland, Ivar} % entries for the author index
% \index{Temam, Roger}  % of the whole volume
% \index{Dean, Jeffrey}

\begin{abstract}        % give a summary of your paper
This paper puts forward a broad critical analysis of the concept of physical time. Clock effect is conceived as a consequence of the variation of the gravitational or pseudo gravitational potential, and it is remarked that only some real clocks measure durations in agreement with the predictions of general relativity. A probable disagreement is expected between radioactive and atomic clocks, in the light of Rovelli's thermal time hypothesis. According to the recent contributions by Rugh and Zinkernagel, the relationship between time and clocks is investigated in order to found on a physical basis the concept of cosmic time. In the conclusive section is argued the impossibility of reducing thermal time to relativistic time.

%                         please supply keywords within your abstract
\end{abstract}
\section{Introduction}
The following points resume the main concepts: 
\begin{description}
\item [1)]analysis of the operational definition of time in Newton and Einstein; \\
\item [2)]distinction between time dilation effect and clock effect in Einstein's theory; \\
\item [3)]considerations about the nonequivalence between atomic clocks and radioactive clocks in the light of the \textit {clock hypothesis}; \\
\item [4)]remarks about the existence, in nature, of thermal clocks that do not behave as relativistic clocks, and about the consequent nonequivalence between thermal and relativistic time, in the light of the \textit {thermal time hypothesis}; \\
\item [5)]reflections about the operational definition of cosmic time in relation to the physical processes, then to the clocks that can be used for measuring it; \\
\item [6)]conclusive remarks about the need of a revision and a consequent refoundation of the concept of time in physics. 
\end{description}
The argumentation, developed in a dialectic ideal debate with experts in the field as Malament, Maudlin, Rovelli and Brown, leads to the conclusion that the problem of time must necessarily be faced starting from the measuring instruments, while too often it was decided to solve it in the dispute between antagonistic theories. The concluding remarks intend to develop the epistemological consequences of the probable existence of different physical times, which is configured as potentially revolutionary.
%%%%%%%%%%%%%%%%%%%%%%%%%%%%%%%%%%%%%%%%%%%%%%%%%%%%%%%%%%%%%%%%%%%%%%%%%%%%%%%%%%%%%%%%%%%%%%%%%%%
\section{The concept of time in Newton and Einstein}
In Newton's mechanics time is a mathematical variable to which it is useful to refer the variation of physical quantities. The usefulness of a variable, however, does not imply that it refers to something real: abstractions, being ideal, do not flow. In Newton's universe bodies move, orbit, attract, internally change, in a becoming in which past, present and future are common to everything and everybody. Simultaneity and durations of phenomena are absolute: the life of a radioactive particle is a time interval between two mathematical instants measured by a universal clock. According to Newton a duration is an abstract property of the whole, whereas the bodies transform as accidents of a substance that, in its parts, changes and moves, but without the need to quantify, if not ideally, its becoming. Space is the place of movements and evolutions of bodies marked by instants and durations extrinsic to them: clocks can only simulate the flow of time by the means of motion. \\Compared to Newtonian mechanics, Einsteinian relativity engenders a radical revolution. With regard to special relativity, the new concept of time is a consequence of the postulate of invariance of the speed of light, from which follows the relativity of simultaneity: two inertial observers in different states of motion do not agree on the instants that correspond to the events that delimit a time interval. In the theoretical framework of special relativity instants and durations are dependent on the observer and the physical universe is spacetime, a mathematical space devoid of matter and energy that we can imagine continuously filled with ideal clocks. These clocks are all synchronized with respect to a given observer, that measures the duration of a phenomenon (for example, the mean lifetime of an unstable particle) through two clocks located in the places where the particle respectively is born and decayed. Since another inertial observer, in motion with respect to the first one, does not agree on the spacetime coordinates of such events and on their respective differences, the phenomenon has a different duration with respect to them: the duration of the mean lifetime of a radioactive substance is therefore not absolute. \\These considerations require a critical analysis of the operational meaning of the concept of duration in Einstein's theory. 
%%%%%%%%%%%%%%%%%%%%%%%%%%%%%%%%%%%%%%%%%%%%%%%%%%%%%%%%%%%%%%%%%%%%%%%%%%%%%%%%
\section{The spacetime of special relativity as space filled with pointsized clocks}
From Einstein's foundational work \cite {ein1} it is deduced that the physical meaning of Lorentz transformations\footnote {Coordinate transformations equivalent to Lorentz transformations have been obtained, by Cattaneo \cite {cat} and Pal \cite {pal}, without the \textit {c}-invariance postulate. According to these authors the limit and invariant speed of light turns out to be a consequence of the principles of homogeneity of time and space and of the principle of causality, so the Einstein choice to put this speed equal to the speed of light in vacuum is based on the assumption of validity of Maxwell's equations, that can therefore be considered the invisible foundations of relativistic theory.} consists in the different quantification of instants and spatial coordinates provided by inertial observers in different states of motion, performing measurements in a universe made of mathematical points, in each of which an ideal clock records the passage of time, after being properly synchronized\footnote {Two clocks are synchronized by sending a light signal from clock A which is reflected by clock B, and then is received back by A. Respectively indicated with $t_a$, $t_b$ and $t_c$ the events corresponding to sending, reflection and return of the signal, according to Reichenbach \cite {rei} must be $t_b= t_a +  \epsilon(t_c  -  t_a)$, where $0< \epsilon <1$. Since only the mean round-trip speed of light can be measured, in special relativity conventionally it is assumed that space is isotropic, so that the one-way speed of light is the same in all directions, and therefore, in agreement with Einstein's convention, $\epsilon= \frac{1}{2}$, then $t_b= t_a +  \frac{1}{2} (t_c  -  t_a)$. This assumption of isotropy, however, has no foundation in general relativity. Whereas in special relativity two clocks B and C, synchronized with a clock A, are synchronized with each other (transitivity of synchronization), the same property does not hold in general relativity. Basri \cite {bas} argues that, since Lorentz contraction and time dilation depend on the procedure of synchronization of clocks, such effects have received, in the scientific literature in matter of relativistic theory, an excessive emphasis, while the clock effect, based on the comparison of proper durations (independent of the procedure of synchronization), measured by real clocks along two nonequivalent world lines between two fixed extreme events, plays, within the Einstein theory, a very important theoretical and empirical role.} with all the others. Given two events that mark the beginning and the end of a phenomenon, the spacetime coordinates that identify them are different with respect to two inertial observers O and O' in relative motion. Assuming that the initial and final events occur, according to O', at the instants $t'_{1}$ and $t'_{2}$ at the same place, so that $x'_{1}$  = $x'_{2}$, such events will not occur at the same place with respect to O, which will assign to them different temporal ($t_{1}$ and $t_{2}$) and spatial ($x_{1}$ and  $x_{2}$) coordinates. Starting from the Lorentz transformation of times: 
 \begin{eqnarray} 
t=(t'+\frac{vx'}{c^2})\gamma
\end{eqnarray}
where $\gamma = \frac{1}{\sqrt{1- \frac{v^2}{c^2}}}$, we obtain:
 \begin{eqnarray} 
t_2 - t_1 =(t'_2 - t'_1)\gamma
\end{eqnarray}
Though apparently it shows a relationship between durations, law (2) actually expresses a relationship between differences of instants, that are postulated to be durations. \\It is necessary to emphasize and deepen the meaning of this seemingly innocuous concept. \\In special relativity the spacetime continuum is a set of mathematical points that correspond, for each observer, to a set of (ideally infinite) synchronized clocks. Every clock of the set provides the information about the instant corresponding to a particular event that occurs at the point where it is placed. Unlike Newton's theory, in which there is only a universal clock that marks the absolute time, whose instants are the same for each observer, in Einstein's theory every observer fills the spatial continuum with a set of synchronized identical clocks, all beating time with the same rate: since the different sets of clocks, associated to observers in different states of motion, are not synchronized between them, two observers in relative motion do not agree on the instant they assign to a given event, so they measure different durations in correspondence to a given phenomenon, which could be the life of an unstable particle. \\In order to compare the measures of the mean lifetime, performed on the same sample by two observers O and O' in relative motion, we generally proceed as follows. \\Let us consider an unstable particle, generated at a given point in space, that decays after a time $\tau_{1}$ measured (through a real clock) by an observer O' at rest with respect to it. If a second observer O is moving at speed \textit{v} with respect to O', he will detect that the birth and the decay take place at different points, between which he measures, with a metric rope, the distance \textit{l}. Known the value of the relative speed \textit{v}, the mean lifetime measured by O will be $\tau_{2} =  \frac{l}{v} $. The two lifetimes show to be different and their ratio $\frac{\tau_{2}}{\tau_{1}}$, in accordance with (2), results to be equal to $\gamma$. In all the experiments on radioactive particles performed in the last century were in fact compared the mean lifetime $\tau_{1}$, measured by an observer O' on a beam of particles at rest, and the mean lifetime $\tau_{2}$ measured on the same beam in flight by the observer O, and it was argued that the mean proper lifetime, measured on a particle that is born and decayed at the same point, is shorter than every mean lifetime measured in flight (whose value depends on the relative speed) by an observer according to which the particle is born and decayed at different points of space. \\In all the experiments (with linear beams or into the CERN storage ring) law (2) was in fact verified using only one real clock, indirectly deducing that to every point of space a given observer associates an ideal clock, so that the clocks of a set are synchronized with each other but not with those of another observer in motion with respect to the first one, and provide a measure of the difference between the instants corresponding to the extreme events that mark the beginning and the end of a physical phenomenon. The infinite clocks of the different sets are therefore virtual idealizations, since in every experiment only one clock was employed, at rest with respect to the particles, to make a direct measurement of their proper duration, while the non proper durations were indirectly measured through the ratio between distance and speed, from which the time dilation law was verified. This law is thus not linked to an effect detected on the different rates of real clocks. \\
These observations lead to the need of a clear distinction between time dilation effect and clock effect, that tests the behaviour of clocks (defined as macroscopic devices for measuring time durations as a function of gravitational and pseudo gravitational potential\footnote {The pseudo gravitational potential is a potential that needs to be introduced by a noninertial observer as a consequence of his acceleration with respect to any inertial observer.}) which,  initially at rest in the same reference frame, are separated along different world lines and at last rejoined. From this distinction some considerations of absolute importance will follow in matter of the operational definition of physical time.           
%%%%%%%%%%%%%%%%%%%%%%%%%%%%%%%%%%%%%%%%%%%%%%%%%%%%%%%%%%%%%%%%%%%%%%%%%%%%%%%%%%%%%%%%%%%%%%%%%%%
\section{Time dilation and clock effect}
From the previous analysis it is deduced that the time dilation effect consists in the different duration of the same phenomenon measured by inertial observers in relative motion, where with respect to one of them (O') the initial and final events occur at the same point of space, while with respect to the other (O) they occur at different points. The effect is symmetrical, in the sense that if O remarks the spatial coincidence of the extreme events, O' measures a dilated time interval if compared to that measured by O. This effect does not test the behaviour of clocks, that in special relativity are assumed to be ideally pointsized and synchronized devices.  \\As claimed by Basri, if you want to broaden the framework of the investigation to general relativity, the clock effect, also known as twin effect, assumes a particular importance. \\In the last century few experiments for testing the clock effect were conceived: the most important were performed by Hafele and Keating \cite {haf}, through four cesium clocks flying on commercial airlines around the Earth (two to the east and two to the west), and by Alley \textit{et al} \cite {all}, through three rubidium clocks in flight along a closed path. In both cases the clocks in flight have measured different durations with respect to those left on the ground, with which they were synchronized at the departure and compared at the return of the voyagers. \\The effect has been theoretically analyzed, in the quoted works, with respect to an inertial reference frame (with the origin at the center of the Earth) and it was considered that, to obtain predictions in agreement with general relativity, it is necessary to describe the phenomenon with respect to any inertial reference frame in which both clocks (on the ground and traveling) are in motion, because the durations measured by clocks are not dependent on the relative speed, but on the speed with respect to the chosen inertial reference frame. It is noteworthy that, if you give priority to the analysis of an inertial observer, you generate a superposition of two effects, one of time dilation (known as speed effect, according to special relativity), and one of gravitational red shift (explained in the theoretical framework of general relativity), thus resulting in a theoretically hybrid process, given that time dilation, as described in the previous paragraph, is an effect linked to the quantification of different durations provided by two inertial observers in relative motion, then conceptually and operationally distinct from clock effect\footnote {With regard to the interpretation of clock effect given by a noninertial observer, we point out the detailed analysis provided by Ashby \cite {ash} of the behaviour, according to a terrestrial observer, of GPS clocks, about which it is necessary to take into account the superposition of three effects: time dilation, gravitational red shift and Sagnac effect.}. \\We therefore believe necessary, particularly in the light of the theoretical analysis given by Basri, but without entering the details of his complex mathematical processing, to explain the clock effect as a simple potential effect, in the following sense. \\We admit, as a necessary consequence of the definition, that a proper time must be measured by an observer in the same state of motion of clock. This operationally means that every clock measures, with respect to a co-moving observer, that therefore can also be noninertial, a proper time linked to its proper period: a proper duration is obtained by multiplying the proper period of the instrument by the number of oscillations of the device that forms its vibrating heart. Since the proper period depends on the gravitational and the pseudo gravitational potential, the measure of the proper duration of a phenomenon, which occurs between two fixed extreme events (departure and return of the plane), according to the co-moving observer will be dependent on the potential in which the clock was during the course of the phenomenon. Since two observers, respectively on the ground and on the plane, each of which takes the measure through his own clock, detect different durations, they will explain the nonequivalent measurements as a simple consequence of the difference of potential. \\The problem then becomes: if two atomic clocks, one on the ground and one on a plane, measure durations in agreement with the theoretical predictions, since the ratio of these durations is in accordance with general relativity, the same ratio will be obtained even through clocks of different construction, for example through radioactive clocks? \\This question hides a problem of deep physical meaning, since the likely nonequivalence of clocks, that will be investigated in the following paragraphs, calls into question the probable existence of times of different nature, of which the relativistic time could be just one of many possible. This suggests that the theory of relativity leaves open the possibility of a plurality of times of which it is not able to predict the intrinsic difference, whose existence could reopen in a traumatic way the debate around this physical quantity, from the analysis of the behaviour of real clocks observed under the same experimental conditions.
%%%%%%%%%%%%%%%%%%%%%%%%%%%%%%%%%%%%%%%%%%%%%%%%%%%%%%%%%%%%%%%%%%%%%%
\section{Minkowski's spacetime}   
In the light of 1905's work on special relativity, the Einsteinian universe is a model of implicit spacetime, in which a three-dimensional continuum (the space) is filled with a set of synchronized clocks, ideally pointsized, that mark the instants corresponding to the spatiotemporal events. \\In 1908 Minkowski formalizes special relativity in terms of a theory of four-dimensional empty spacetime, according to which the duration of a phenomenon is the length of the world line measured by a clock as a proper time. \\Though  apparently it appears only a different mathematical formulation of the same theoretical interpretation of physical phenomena, Minkowski's model takes a conceptually decisive step forward, as it implicitly considers the durations directly measurable by clocks, conceived as real instruments able to quantify the length of the world lines between two extreme events. \\Through the theory of general relativity, Einstein in 1916 develops Minkowski's model by  introducing the effect of the presence of matter-energy that determines, depending on its local density, the curvature, hence the metric of spacetime. \\The following analysis intends to show that general relativity is built on a geometrical model of spacetime in which the measurements of the durations can be obtained only through instruments having a particular internal structure.
%%%%%%%%%%%%%%%%%%%%%%%%%%%%%%%%%%%%%%%%%%%%%%%%%%%%%%%%%%%%%%%%%%%%%%
\section{The measure of durations in general relativity }    
While according to Newton's theory real clocks provide a relative, approximate\footnote {The measure of durations in Newton is relative and approximate as obtained by devices that simulate the flow of absolute time, by definition mathematic, then ideal.} and external measure of the absolute duration, therefore of the variable \textit {t}, according to  Einstein's theory clocks measure the length $\tau$ of the world line, hence not \textit {t}, which appears to be a simple mathematical label without physical meaning. In general relativity the evolution of bodies and phenomena is not a function of an independent and preferential variable as, instead, it happens with Newtonian time, that plays the role of the independent parameter to which every evolution is referred. \\According to Rovelli \cite {rov1} general relativity describes the evolution of observable quantities relative to one another, without conceiving one of them as independent: in general relativity there is a potentially infinite number of variables, each of which, in turn, conventionally can assume the role of time variable, without having to admit the objective existence of a physical quantity called time. \\We point out that in Rovelli's analysis, albeit indisputable in terms of formal coherence and mathematical rigor, the evolution is limited to the variation of the relative position of bodies, that certainly can be interpreted in relational sense, without having to explicitly introduce a time variable. Rovelli, in essence, suggests to ignore a quantity that, in mechanics, has always played only a conventional role (absolute or relative, not matter), as reducible, both in Newton and in Einstein, to a useful mathematical variable. While Newton explicitly defines the absolute time a mathematical abstraction (the duration) that we derive from motion, the mathematical nature of relativistic durations, measured by clocks along the world lines they describe, is in fact also present in the theoretical framework of general relativity, because every clock, when it describes nonequivalent world lines between two fixed extreme events, records different durations only if it is conceived as an ideal clock, in the light of the \textit {clock hypothesis}. \\In matter of operational definition of relativistic time, some recent reflections by Brown are remarkable. In agreement with a Bell's work \cite {bel} and with an observation by Pauli \cite {pau}, Brown notes \cite {bro} that clocks do not measure time as, for example, the thermometers measure the temperature or the ammeters measure the electric current: their behaviour correlates with some aspects of spacetime, but not in the sense that spacetime acts on them in the way a heat bath acts on a thermometer or a quantum system acts on a measuring device. We wonder: to which time do the clocks in general relativity refer? Which is the meaning of the measurement of a quantity that, in the light of the above Rovelli's observations, is only mathematically useful and could be forgotten? In fact, relativistic clocks record different durations between two fixed extreme events only if, during a given phenomenon, they are in different gravitational or pseudo gravitational potentials, so they must have special requirements to provide measures in accordance with the theory. This fact clearly does not mean that relativity exhausts inside it every possible operational definition of time. The systematic and experimental doubts raised by Brown about the possibility that some quantum clocks\footnote {Brown refers in particular to Knox \cite {kno}, who critically discusses an essay by Ahluwalia \cite {ahl}, in which the author suggests that some "flavour-oscillation clocks" might behave in a way that threatens the geometricity of general relativity. Ahluwalia shows that these quantum mechanical clocks do not always redshift identically when moved from the gravitational environment of a non–rotating source to the field of a rotating source and claims that the non-geometric contributions to the redshifts may be interpreted as quantum mechanically induced fluctuations over a geometric structure of spacetime. Knox instead argues that the purported "incompleteness-establishing" is not at odds with any of general relativity foundations.} do not behave as relativistic clocks, as they may not be in accordance with the clock hypothesis, in addition to the important objection about the fact that relativistic clocks do not measure in the classical sense, since they merely record the correlation between the instrument and particular aspects of the spacetime structure, implicitly open the door to the recognition of a potential multiple reality of time, that requires new conceptual and operational tools to be explored.
%%%%%%%%%%%%%%%%%%%%%%%%%%%%%%%%%%%%%%%%%%%%%%%%%%%%%%%%%%%%%%%%%%%%%%
\section{Clock hypothesis}   
Although it is now widely agreed that the experimental tests carried out in the last century have provided a virtually definitive proof of relativistic laws, we believe necessary to interpret the conclusions of the crucial experiments by which the relativistic theory of time has been confirmed. For this purpose we will refer to the recent treatises by Malament \cite {mal} and Maudlin \cite {mau}, that reconstruct the relativistic building from simple theoretical considerations of geometrical nature. We wonder: what do relativistic clocks measure, according to these authors? Malament defines ideal clock a pointsized object\footnote {Malament's definition is in agreement with Kostro's one \cite {kos}, which remarks that relativistic clocks should be virtually point-like, whereby every real clock, having an extension, will always be far from the ideal model with which it should be compared.} that records the passage of time along a world line:
\begin {quote} 
One might construe an “ideal clock” as a pointsized test object that perfectly records the passage of proper time along its worldline, and then ... assert that real clocks are, under appropriate conditions, to varying degrees of accuracy, approximately ideal. 
\end {quote}   
According to Malament, Maudlin gives the following definition of ideal clock:  
\begin {quote} 
An ideal clock is some observable physical device by means of which numbers can be assigned to events on the device's world-line, such that the ratios of differences in the numbers are proportional to the ratios of interval lengths of segments of the world-line that have those events as endpoints. 
\end {quote} 
We believe that the possibility that real clocks approximately behave as ideal clocks constitutes the most critical point of spacetime theory. It is not clear, in fact, if Malament intends to refer to elementary particles as implicit clocks, able to record the flow of time along a world line: if it were possible, it should be identified the evolutionary process that allows an elementary particle to get a measure of duration related to a phenomenon that occurs inside it. \\A real clock, to be such, must allow to measure a duration either through a periodic phenomenon or through an internal change, as the radioactive decay could be. If, as it can be deduced from Maudlin's definition, an experimental proof of clock effect requires that the ratio between the measures provided by two clocks of identical construction, describing two nonequivalent world lines between the same extreme points-events \textit {p} and \textit {q}, is equal to the ratio between the measurements provided by any other pair of clocks, between them equivalent but different from the previous ones, the relativistic theory of time cannot obviously be tested only with ideal clocks. While Malament suggests, without investigating it, the hypothesis of pointsized clocks, Maudlin, taking up an Einstein's idea, limits to consider the behaviour of light clocks. Relativistic theory, according to these authors, is built on simple as clear purely geometric assumptions and takes the form of a theory of spacetime and of time-light, in which the behaviour of real clocks is not investigated. Malament remarks that the apparent clock-twin paradox is born from the ignorance about the different behaviour of clocks in free fall (therefore along a geodesic line) or subjected to acceleration, and he solves the problem within the geometric model of spacetime, through a logically irrefutable argumentation. According to Malament the misunderstanding arises because, mistakenly, someone can believe that the theory of relativity does not distinguish between accelerated motion and free fall: Malament's solution, shared by Maudlin, is to recognize that the two situations are not symmetrical and that the two clocks must quantify different durations measured as lengths of the world lines by them described (\textit {clock hypothesis}). Einstein's theory, interpreted by these authors as a philosophy of spacetime, is founded on a model of universe as a four-dimensional perfect fluid: time, according to this philosophy, is implicitly contained in the world lines, regardless of the real clock that quantifies the duration. \\We renew, in this theoretical context, the innocuous as immediate objection previously raised: since real clocks do not meet the definition of ideal clock given by Maudlin and Malament, can they at least be considered an empirical approximation of this definition? \\In order to specify in more detail this objection, we propose the following two problematic issues, respectively related to the speed effect and to the gravitational red shift effect. \\
Firstly, let us consider two clocks of identical construction that, initially at relative rest and synchronized, describe two nonequivalent world lines, on rectilinear paths, due to the different speed gained, after a short acceleration, by one of them with respect to the other: after having described different segments of straight line, the clocks rejoin due to the slowdown of the faster one. Since their proper period must be the same\footnote {The periods are different only in presence of a gravitational or pseudo gravitational potential difference.}, in order to experimentally verify if indeed they have measured different durations, as claimed by Maudlin and Malament in agreement with Einstein's theory, it is necessary to count the number of beats scanned along the paths traveled in uniform rectilinear motion. Since in the experiments carried out so far, in which the effect was tested with clocks in relative motion (see the above mentioned performed by Hafele and Keating and by Alley \textit {et al}), the clocks in flight did not describe linear paths, we believe that an experimental evidence of different measures of duration due to the only speed effect on both rectilinear trajectories is not to date available. A direct verification of this seemingly simple experimental fact has undoubtedly an important meaning, since it would prove or less whether, as consistently argued by Maudlin and Malament, the clock effect on a straight line (being negligible accelerations and decelerations of the one that moves away and then rejoins with its twin) is a simple consequence of the different lengths of the world lines described by clocks, so that the one that has traveled the shortest line must have scanned, without changing its proper period, a smaller number of beats. \\
Secondly, let us consider two clocks that describe nonequivalent world lines due to the different gravitational potential, for example two clocks that, initially synchronized in the same laboratory, are separated so that one remains at the sea level and the other one is led on a high mountain, and finally they are rejoined in the laboratory and to one another compared at relative rest (as in the Briatore and Leschiutta experiment \cite {bri}, in which the effect was tested on a pair of atomic clocks). Assuming that two pairs of clocks of different construction are observed in the above described conditions, we wonder if the ratio between the durations measured by the first pair, for example of atomic clocks, on the world lines 1 and 2, will be equal to the ratio between the corresponding durations measured on the same lines by a second pair, for example of radioactive clocks. \\
The following objection, contrary to the two above raised, is predictable: is it not a permanently acquired experimental fact that the synchronization of the GPS atomic clocks is rigorously explained by Einstein's theory?\\ It is remarkable that in Ashby's theoretical analysis the fact that clocks in orbit describe curvilinear trajectories was not carefully considered, whereby their proper period is different due to the different potential, both gravitational and pseudo gravitational, compared to that of clocks on the ground. Since it is possible to explain the different measures of clocks in the light of an easier as well as theoretically more elegant potential effect, the following conceptual and operational problem arises. \\Though it is indisputable that the synchronization of the GPS clocks was made in agreement with the theory of relativity, of whose logical-mathematical consistency it today provides one of the most convincing and decisive proofs, the above critical objections do not put into question the predictive ability of Einstein's theory about the operation of particular categories of clocks, but the generalizations in matter of measure of durations that have been implicitly deduced. Because the GPS is a system of synchronized atomic clocks, we are not authorized to foresee that clocks having different internal structure must behave in the same manner as the limited category of real clocks so far employed in experiments performed to test the clock effect. It follows that, if from the analysis of the operation of different real clocks a blatant inconsistency arises with respect to the clock hypothesis on which, according to Maudlin and Malament, the theory has been founded, a refoundation of the problem of time in physics would be necessary. It is what we propose to do in the following sections. 
%%%%%%%%%%%%%%%%%%%%%%%%%%%%%%%%%%%%%%%%%%%%%%%%%%%%%%%%%%%%%%%%%%%%%%
\section{Real clocks}   
The experimental devices employed for measuring time intervals, distinguished on the basis of their internal structure, are largely attributable to the following categories:
\begin {itemize} \renewcommand{\labelitemi}{\normalfont }
\item \textbf {gravitational clocks}, in which the measurement of a duration can be obtained: a) indirectly, by exploiting a time scale deduced from the positions of particular celestial bodies (ephemeris time); b) directly, by exploiting the motion, approximately harmonic for small oscillations, of a pendulum, which can be considered, together with the hourglass, one of the most simple but effective gravitational clocks;\\
\item \textbf {balance-wheel clocks}, whose mechanism is constituted by a swinging wheel, rotating around an axis, by a spiral spring that develops an elastic force and by an anchor (the system said escapement) that gives the wheel small thrusts at the right time: in principle, the phenomenon is equivalent to the linear harmonic oscillation of a mass-spring system, whose proper period depends only on the mass and on the elastic constant of spring; \\
\item \textbf {quartz clocks}, in which the period is determined by the oscillations of a quartz crystal: a quartz resonator, under the electronic point of view, is equivalent to an RLC circuit, with a very precise and stable resonance frequency;\\
\item \textbf {atomic clocks}, whose proper period is directly linked to the energy difference between two fixed quantum states in an atom; \\
\item \textbf {radioactive clocks} \cite {bor1}, in which the decay produces a quantity of daughter substance from a given amount of initial parent substance: it should be specified that, while gravitational, atomic, quartz and balance-wheel clocks refer to phenomena involving the alternation of identical phases, so that the measure is related to the count of the number of oscillations and of its submultiples, in a radioactive clock the measure of a duration implies, given the initial amount of parent substance, the quantification of the decayed mass in correspondence to the world line described by the sample between two fixed extreme events. 
\end{itemize}

\section{Atomic and mechanical clocks}   
As accurately remarked in previous works \cite {bor2,bor3}, from the relationship between the metric coefficient $g_{00}$ associated to the coordinate \textit {ct} and the potential $\varphi$ in a static and weak gravitational or pseudo gravitational field, the following approximate relationship between the proper period \textit {T} of a relativistic clock and the potential $\varphi$ can be deduced, $T_{0}$ being the period outside the field:
\begin{eqnarray} 
T(\varphi)=T_0 (1 - \frac{\varphi}{c^2})
\end{eqnarray}
This law was experimentally tested on atomic, optical, quartz, based on the use of maser or at resonant cavity clocks. \\
As regards atomic clocks, since the atom of a given substance can assume only particular excited states (characteristic of the element to which it belongs), when passing from a first energy level to a lower second one the atom emits a given amount of energy, or it absorbs the same amount when it passes from the lower to the superior level. This transition is thus linked to the emission or to the absorption of an electromagnetic radiation of energy $\Delta E$, linked to the corresponding frequency through the relationship $\Delta E=h\nu$. Atomic clocks provide measures in agreement with relativistic theory because the energy difference $\Delta E$ between two fixed energy levels in an atom depends on the gravitational or pseudo gravitational potential $\varphi$ according to the relationship $\Delta E={\Delta E_0}{\sqrt{1+2\frac{\varphi}{c^2}}}$, where $\Delta E_0$ is the energy difference in a null potential. Since the proper period of an atomic clock is related to the energy difference according to the relationship $T = \frac{h}{\Delta E}$, you get that $T$ is approximately in agreement with (3). In such devices the duration quantified along different world lines is therefore a consequence of the variation of the proper period of the instrument as dependent on the energy emitted in correspondence to the given quantum transition, which in turn depends on the gravitational or pseudo gravitational potential. \\ In a recent work \cite{bor3} a theoretical analysis has been proposed about mechanical clocks, that in all probability, at least limiting to pendulums and spring clocks, do not behave as relativistic clocks. As in particular regards pendulums, they behave in clear disagreement with the predictions, since, going up in altitude (increasing the gravitational potential), they reduce their proper frequency of oscillation, while, according to (3), they should increase it, as indeed cesium clocks do. Einstein  himself, in a footnote on his 1905's work, discarded the possibility of using pendulum clocks, as devices that do not work at zero gravity (for example, in free fall), since the oscillations are present only in gravitational or acceleration fields.  \\ It is fundamental to remark that a relativistic clock requires an internal process that does not cease to occur in free fall. \\ In the next section we will limit to the comparison, that we believe paradigmatic, between atomic and radioactive clocks.
 %%%%%%%%%%%%%%%%%%%%%%%%%%%%%%%%%%%%%%%%%%%%%%%%%%%%%%%%%%%%%%%%%%%%%%
\section{Radioactive clocks. Nonequivalence between radioactive and atomic clocks}   
As previously remembered, the experiments on unstable particles in flight have proven that the mean lifetime indirectly measured (by calculating the ratio between the mean distance, traveled before the disintegration, and the speed) when muons are observed in flight (birth and decay occur at different points) is dilated if compared to the measurement directly made in the laboratory (in which emission and decay occur at the same point) on a sample of muons at rest. Is noteworthy, as regards the behaviour of unstable particles in flight, the experiment with muons shot at a speed close to \textit {c} into the storage ring at CERN \cite{bai}: the element of novelty is in this case the presence of a centripetal acceleration of the order of $10^{18} g$, and therefore of a huge pseudo gravitational potential, with respect to the laboratory reference frame, whose origin is at the center of the ring. Since this potential did not produce different effects if compared to those observed in the experiments with muons in linear flight in the atmosphere, some important considerations can be inferred in relation to the possibility of considering clock a sample of unstable particles. \\In order to remark the distinction between time dilation and clock effect, we point out the theoretical analysis by Eisele \cite {eis}, that used perturbation techniques in the theory of the weak interaction to calculate the approximate lifetime of muons, and concluded that the correction to the calculation based on the clock hypothesis for accelerations of $10^{18} g$ would be less than 1 part in $10^{25}$, many orders of magnitude less than the accuracy of the 1977 experiment. \\According to Eisele the clock hypothesis says that in nature there are ideal clocks with the property that their timekeeping is independent of their acceleration or at least does not depend on it in a measurable way, hence they should always show their proper time $\tau = \int_0^t \sqrt{1-v(t')^2} \, dt' $. In order to verify whether such ideal clock really exists, Eisele considers that an unstable elementary particle can be used as a clock because of its characteristic mean lifetime. His articulate demonstration leads to conclude that the time-keeping of this kind of clock depends essentially on its energy, so that 
\begin{eqnarray} 
\Delta t = \frac{E}{E_0} \Delta \tau
\end{eqnarray}                                                                                                  where $\Delta t$ and $\Delta\tau$ are respectively the non proper and the proper time of the clock, $E_0$ its rest energy in a field-free space and \textit {E} its total energy, independently of whether the increase of energy comes from a kinetic energy or from a "zero-point-energy" in a magnetic field\footnote {Eisele properly notes that in the case of (negative) gravitational energy because of the gravitational redshift exactly the opposite result is true: $\Delta t = \frac{E_0}{E}\Delta \tau$}. \\The remarkable fact is that, being $\frac{E}{E_0}  = \gamma$, the dynamical relationship (4) between $\Delta$\textit {t} and $\Delta\tau$ is equivalent to the kinematical relationship (2), and therefore it is independent of the device. In fact, the special relativistic effect of time dilation implicitly refers to time as a quantity external to clocks conceived as ideal devices, that flows at different rates in reference frames in relative motion. In this sense, according to Brown, ideal clocks do not measure anything inside them and the countless experiments performed in the last century on time dilation do not say anything about the internal time of clocks, since, according to Will \cite {wil}, "the proper time between two events is characteristic of spacetime and of the location of the events, not of the clocks used to measure it"\footnote {At this regard, it is curious that Einstein himself had recognized very late, in 1949 \cite {ein2}, the "sin" of treating, in his 1905's paper, rods and clocks as primitive entities, and not as "moving atomic configurations" subject to dynamical analysis.}. \\The question moves necessarily to real clocks and to clock effect.  \\Being a radioactive clock a device that measures a proper duration through the amount of decayed substance in relation to a given phenomenon (for example, the duration of the trip of the plane that carries the same radioactive sample), we may wonder if such a device, that certainly also works in free fall, behaves as a relativistic clock. The fact that an acceleration of the order of $10^{18} g$ did not alter the internal structure of the particles allows to deduce that the pseudo gravitational potential, proportional to this acceleration, with respect to an observer in the laboratory, did not influence the proper period of clocks, producing experimental measurements identical to those obtained on linear beams. This implies that radioactive clocks are independent, in the light of the principle of equivalence, of gravitational and pseudo gravitational potential and therefore they are nonequivalent to atomic clocks, whose proper period is dependent on the above mentioned potential.    
In papers written by leading experts in the field it is taken for granted that atomic and muon clocks behave in accordance with the relativistic predictions, superposing, in particular in the case of Brown in the quoted analysis of the behaviour of clocks (and rods) in general relativity, the effects of speed and gravitational red shift. The anomalous fact that a curvilinear motion, in presence of an high acceleration, has produced on muons only a speed effect is interpreted by Brown as a clear agreement with the clock hypothesis, then as a proof that  acceleration does not have any effect on clocks. \\The important fact, that Brown does not consider, is that a corotating observer, at relative rest with respect to the muons, must deduce that the centrifugal potential $V_c = -\frac{1}{2}\Omega^2r^2$, \textit {r} being the radius of the circumference described by the muons and $\Omega$ the angular speed, has no effect on the measurement. \\Let us make the hypothesis to bring a radioactive clock on a plane and to compare it with a radioactive clock on earth. We know that the durations measured by atomic clocks are a function of the gravitational potential or, in the case of the experiments performed by Hafele and Keating and by Alley \textit {et al} (in which the different gravitational potential of clocks in flight with respect to those on the ground has produced negligible effects), of the pseudo gravitational (centrifugal) potential. It is important to remember that in the quoted experiments the planes have described curvilinear trajectories, whereby with respect to corotating observers the atomic clocks were in a centrifugal potential that has significantly altered their proper period, if compared to those on the ground. Since the pseudo gravitational potential has no effect on the proper period of radioactive clocks, it is easy to deduce that their proper mean lifetime during the trip would not have  been altered. At this regard, we believe necessary to control if: \\ \\a) in an experiment like those performed by Hafele and Keating or by Alley \textit {et al}, the ratio between the decayed substance in initially identical radioactive samples on earth and on a plane is equivalent to the ratio between the durations measured in the same conditions by atomic clocks; \\ \\b) while two atomic clocks, initially synchronized in the same laboratory and separated bringing one of them at high altitude, measure different durations if rejoined and compared in the same laboratory, the same  phenomenon happens with a pair of radioactive clocks. \\ \\The issue in fact should not be focused on the possible influence of acceleration in experiments whose goal is the proof of the time dilation effect, then of the relationship between non proper and proper time, but on the experimental test of the clock effect, then of the relationship between the proper times of two clocks that, at relative rest at the beginning, move away from each other and remain in different gravitational or pseudo gravitational potentials, and finally rejoin in the same reference frame. \\Besides the remarked differences about the behaviour in relation to the potential, atomic and muons clocks are also radically different in relation to the specific kind of internal transformation: reversible in atomic clocks, irreversible in muon clocks. If an atomic clock after a trip returns at the point of departure, the measured duration is not linked to an amount of transformation that has influenced its internal state, that returns identical to the initial one, while in a radioactive clock the amount of decayed substance is linked to an irreversible phenomenon that prevents the spontaneous restoration of its initial state. This important differentiation about the internal phenomenon that leads to measure time through clocks is completely neglected in the framework of Einsteinian relativity and in all the theoretical works inspired by the Einsteinian theory. This problem was not detected by Brown, according to which the muons are implicit clocks able to measure the length of the world lines, while he remarks the open question about the possible existence of  clocks, as the above mentioned Ahluwalia's flavor-oscillation clocks, whose behaviour could be not in agreement with relativistic predictions, namely with the clock hypothesis. \\The issue, in the light of the analysis presented in this work, is far more radical of the possible existence of some experimental anomaly that, according to a more careful analysis, as noted by Knox, could be not in disagreement with Einstein's theory. We believe, therefore, that the different behaviour of atomic and radioactive clocks implies the possible existence in nature of different classes of clocks, among which only a little number belongs to the implicit class of devices that can be defined, in the light of law (3), relativistic clocks. Since we believe impossible to investigate the nature of time regardless of clocks that are used to measure durations, the existence in nature of clocks whose proper period is not in agreement with relativistic predictions has the value of an experimental and theoretical discovery, since it implies the existence of times of different nature.
%%%%%%%%%%%%%%%%%%%%%%%%%%%%%%%%%%%%%%%%%%%%%%%%%%%%%%%%%%%%%%%%%%%%%%%%%%%%%%%%%%%%%%%%%%%%%%%%%%
\section{A different dimension of time}
The above analysis leads to deduce that: 
\begin{description}
\item [1)] a Newtonian clock is a useful device that can only simulate the measure of the absolute time, by definition mathematical, then ideal; \\
\item [2)] in special relativity every observer should use a set of infinite synchronized clocks, but in every experiment to verify the time dilation effect only one real clock has been employed, at rest with respect to the particles, to obtain a direct measurement of the proper duration, while the non proper durations have been indirectly measured through the ratio between distance and speed; \\
\item [3)]in general relativity a clock measures a duration as length of a world line only if its proper period is linked to the gravitational and the pseudo gravitational potential in agreement with law (3): only some real clocks operate in agreement with this law. 
\end{description}
For completeness, we ask if and how the concept of time has been theoretically and operationally explored in quantum theories. Barbour suggests \cite {bar} that in quantum mechanics the physical reality can be interpreted as a set of snapshots without actual evolution. According to  Rovelli \cite {rov2}, quantum gravity reduces the fundamental reality of phenomena to a network of relations between quantum covariant fields, in which the reality of time seems to be illusory. In quantum theories time is therefore considered unreal and obviously clocks are useless. \\The fact  that in all the fundamental theories the concept of irreversibility of the evolution, that real clocks record when their measures are linked to irreversible phenomena, is neglected, clearly does not mean that time is unreal, but only that these theories do not refer to an operational concept of time linked to irreversible transformations. In this optics it is fundamental the distinction between the internal evolution of bodies and the evolution linked to the variation of their relative position as it is described in mechanics: the true operational essence of physical time must in fact be searched in thermodynamics, where it is generated inside the measuring instruments as a necessary product of an irreversible evolution. The description of physical phenomena provided by thermodynamics seems implicitly to reveal the existence of a dimension of time different from the one that emerges from the fundamental theories: the following hypothesis, formulated by Rovelli, is the possible foundation of a new conceptual exploration of real phenomena and of a new operational definition of time.%%%%%%%%%%%%%%%%%%%%%%%%%%%%%%%%%%%%%%%%%%%%%%%%%%%%%%%%%%%%%%%%%%%%%%%%%%%%%%%%%%%%%%%%%%%%%%%%%%
\section{Thermal time hypothesis. Nonequivalence between thermodynamic and relativistic time}
 Rovelli states \cite {rov1} that in statistical mechanics it is possible to introduce the \textit {thermal time hypothesis}, according to which, though in nature there is not a preferential time variable \textit {t} and a state of preferential equilibrium a priori identifiable does not exist, since all the variables are on the same level, if a system is in a given state $\rho$ it is statistically possible to identify, from this state, a variable $t_\rho$ named thermal time\footnote {The following is the original Rovelli's formulation: "In nature, there is no preferred physical time variable \textit {t}. There are no equilibrium states $\rho_0$ preferred a priori. Rather, all variables are equivalent; we can find the system in an arbitrary state $\rho$; if the system is in a state $\rho$, then a preferred variable is singled out by the state of the system. This variable is what we call time".}. Calling time a certain variable\footnote {According to Rovelli, the thermal time variable $t_\rho$ appears to be the parameter of the flow of the quantity $H_\rho$ (thermal Hamiltonian) defined by the equation $H_\rho=-ln\rho$ and is defined thermal clock any device whose reading linearly increases as a function of $t_\rho$ .}, says Rovelli, we are not making a statement concerning the fundamental structure of reality, but a statement about the statistical distribution used to describe the macroscopic properties of the system under observation. What we empirically call time is therefore the thermal time of the statistical state in which a system is observed, when it is described as a function of the macroscopic parameters we have chosen. Time is thus the expression of our ignorance of the microstates, a conceptual simplification arising from a high number of variables that chaotically change. In thermodynamics comes to light, in its irreducible nature, the reality of becoming intrinsic to bodies and systems, on which the Newtonian theory of mechanics and gravitation, the Einsteinian theory of relativity and quantum mechanics have never focused the attention, since these theories merely describe the dynamical evolution as it appears in interactions involving bodies or systems of which we neglect or ignore the internal structure. \\We have seen that radioactive clocks, as clearly demonstrated by the experiment carried out in the storage ring at CERN by Bailey \textit {et al} in 1977, are independent of the centrifugal potential, from which we have derived, in the light of the interpretation of clock effect as a potential effect, that the measurements obtained by atomic clocks and radioactive clocks cannot be equivalent. Since the radioactive decay is a statistical irreversible phenomenon, we believe that radioactive clocks, in accordance with Rovelli's definition, are clear examples of thermal clocks and we explicitly formulate the hypothesis that thermal time has a different nature with respect to relativistic time. \\The theoretical and operational reality of thermal time implies therefore the need of a revision and a consequent refoundation of the concept of time in physics \cite {bor4}.  
%%%%%%%%%%%%%%%%%%%%%%%%%%%%%%%%%%%%%%%%%%%%%%%%%%%%%%%%%%%%%%%%%%%%%%%%%%%%%%%%%%%%%%%%%%%%%%%%%%
\section{Cosmic time}
Since the issue has implications too broad to be thorough here, we provide only a brief mention to a recent work by Rugh and Zinkernagel \cite {rug} about the physical basis of cosmic time. In relation to the Friedmann-Lema\^{i}tre-Robertson-Walker (FLRW) metric, on which the description of the universe at large scales is founded, the authors propose an articulate reflection on the relationship \textit {t} $\leftrightarrow$ \textit {time}, \textit {t} being a theoretical parameter that can be interpreted as time, so on the need to explore its meaning in relation to specific physical processes. The definition of cosmic time must be linked to standards clocks, therefore to a phenomenon (core of a clock) that allows to build a time scale based on a scale-setting physical process. \\The authors remark, extrapolating the cosmological model towards the origin of spacetime, that such a scale is highly problematic to spot: a first problem arises about the quark-gluon phase transition ($10^{-5}sec$ from the origin) where, as they claim, there might not be bound systems left, and the concept of a physical length scale to a certain extent disappears; a second even more serious problem occurs at $10^{-11}sec$, in correspondence to the electroweak phase transition, where it becomes almost impossible, in the context of the current knowledge, to identify a physical basis for the concepts of length, energy and temperature scale, that in modern cosmology are closely linked to the concept of time. \\In the light of Rugh and Zinkernagel's analysis, related to the implicit belief about the alleged existence of an operational meaning of time in the theory of general relativity that, through the FLRW model, constitutes the foundation of modern cosmology, we believe that the problem of the nature of cosmic time can be summarized in the following questions:
\begin{description}
\item [1)] in which sense is it reasonable to consider the universe an evolving system?\\ 
\item [2)] is this evolution linked to a global movement or to an internal transformation?
\end{description}
These questions at first sight may seem naive or of little interest, since the modern cosmology takes the universal evolution for experimentally granted, deducing from it that the origin of the universe is a spatiotemporal singularity to which we refer the beginning of spacetime, then the origin of cosmic time. \\At this regard, we wonder: if cosmic time implies that the universal evolution may be quantified by measuring an amount of expansion, on which physical basis is it theoretically or experimentally possible to obtain this measure? \\The authors believe, at least in the light of the current state of the knowledge, that there is no answer to this question, since the Big Bang model merely postulates a virtual extrapolation backwards in time, assimilating it to a movement. We believe that this movement, in hindsight, has no theoretical or empirical basis, because the concept of motion involves the concept of space as existing entity on which its operational definition must be founded, and in this case there is no external space available for the expansion. In fact the theory predicts that the expansion happens creating time and space: the space of motion should be created along with the motion, and cosmic time should be related to the speed at which the expansion occurs. Since the physical concept of speed implies a pre-existing space and a measurable time to be operationally founded, we can deduce that the evolutionary time of universe is not founded on a sound operational basis, as it cannot be measured through a motion for which a length scale, that allows to make measurements, does not exist. At this regard, in the section "The cosmic scale factor \textit {R} as a clock", Rugh  and  Zinkernagel remark the need of a fixed physical length scale which does not expand, or which expands differently than the universe, and they properly observe: 
\begin {quote}
Indeed, if everything (all constituents) within the universe expands at exactly the same rate as the overall scale factor, then ‘expansion’ is a physically empty concept. (...) If there are no bound systems, and not least if there are no other physically founded fixed length scales, in a contemplated earlier epoch of the universe, is it then meaningful to say that the universe expands? Relative to which length scale is the expansion of the universe to be meaningfully addressed?
\end {quote}
The possibility remains of measuring the cosmic evolution through thermal clocks in the sense of the above Rovelli's definition. \\By investigating the possibility of building clocks on the basis of phenomena of thermodynamic-statistical nature, in which time is related to the temperature of the cosmic background or to the radioactive decay, the authors implicitly open the door to the problem of the nature of time, although they do not propose any explicit reflection on how to deal with it. Besides the conception, that from Aristotle has entered  the classical physics and the Einsteinian relativity, according to which time and movement are two faces of the same medal, it must be recognized a still largely unexplored conception of time related to the transformations of bodies and systems, derived from thermodynamics, whose meaning is conceptually and operationally different from the previous one. \\The question is physical as much as philosophical: not being able to measure the time-movement of the universe expansion for the above reasons, is it possible to estimate the age of the universe by measuring the evolutionary time of a body or a system of bodies inside it? \\Rugh and Zinkernagel suggest to use particular phenomena to build theoretical and experimental models of implicit clocks: the temperature of the cosmic background, the black body radiation, the decay of unstable particles and the atomic and nuclear processes. We believe, however, that in each of these phenomena it is inevitable to create a virtuous circle with which, as the authors point out in the section "Possible types of clocks in cosmology", we are faced when we admit that time exists independently of clocks:
\begin {quote}
In general, when we try to associate the parameter \textit {t} (= \textit {t}(\textit {C})) with some physical process or a physical concept \textit {C}, we cannot require that the concept \textit {C} does not depend on \textit {t}. Indeed, insofar as time is a fundamental concept, there are (virtuous) circles of this kind in any attempt to specify the time concept. However, we shall require that \textit {C} is not just a mathematical concept, but a concept which has a foundation in physical phenomena (in which case we shall say that \textit {C} has a  physical basis). 
\end {quote}
Ultimately, the fundamental problem can be summarized in the alternative between the postulate of a reality of time external to clocks and the need to anchor the operational definition of time to real clocks, whereby different cores of clocks may measure times of physically different nature. This possibility opens a new research perspective about the physical reality of time, taking a step beyond the assumption about its absolute or relative nature that, respectively in classical Newtonian and in relativistic Einsteinian theories, was always conceived as external to the instruments used for measuring the durations.
%%%%%%%%%%%%%%%%%%%%%%%%%%%%%%%%%%%%%%%%%%%%%%%%%%%%%%%%%%%%%%%%%%%%%%%%%
\section{Operational definitions of time and different levels of physical reality}
A critical analysis of the concept of time in the light of the fundamental physical theories leads to deduce that these theories describe different levels of physical reality, that can be divided into: 1) quantum; 2) relativistic; 3) thermodynamic. Since every theory provides an independent description of phenomena on the basis of different principles, we believe not productive to look for links or deeper syntheses that allow to merge irreducible theoretical visions, in order to draw a unitary conceptual framework on which the knowledge of the ultimate structure of reality could be founded. \\As concerns time, a careful analysis leads to the following conclusions. \\While in quantum mechanics the interactions seem to occur instantaneously in an absolute space (entanglement is the experimental proof), in relativity they require a finite time, since a limit and invariant speed exists and fields involve the structure of spacetime, whose curvature allows to explain gravitational phenomena and that supports each other interaction. The reality of relativistic time is linked to the measurements provided by clocks that operate in accordance with law (3), that expresses the relationship between proper period and potential: only the set of electromagnetic clocks (to which belong atomic clocks and other equivalent devices) contains instruments that work in agreement with this law, whereby we have deduced the above explained problems about the internal consistency of theory in matter of measurement of durations. The evolutionary nature of time emerges at a level of description of real phenomena in which the model of spacetime, in turn inherently problematic, is not fundamental: the evolutionary time is measurable as generated by clocks that register durations in agreement with Rovelli's thermal time hypothesis. The models of physical reality therefore do not necessarily imply one another, each one providing an autonomous theoretical description based on a logical-operational structure different from the others: the only indisputable fact is that the evolutionary nature of irreversible time is not reducible to more fundamental theories, in which time itself is not tied to an intrinsic becoming of bodies and clocks. If, therefore, digging toward the quantum level at a first sight the evidence appears of the reality of relativistic spacetime, which at a deeper level of observation (at the level of quantum fields) fades giving way to the reality of a reversible chaos of instantaneous interactions, we are not entitled to conclude that the true physical reality is the one that appears at a relativistic or quantum scale. It is especially necessary to take conscience of the operational reality of irreversible time, that radically revolutionizes the theoretical framework, as if the phenomenal peculiarities that appear at a macroscopic scale were not explainable by the fundamental theories. \\The necessity to radically reinterpret the concept of time in the light of thermal and, more in general, statistical phenomena, teaches that the physical theories do not always have to yield to reductionist impulses. Though the  evolution of physics has been marked by the understanding of empirical phenomena in the light of unitary theoretical principles (for example, electric and magnetic phenomena have found a complete synthesis in Maxwell's theory), nature seems to shy away, in matter of time, the will of reducing the complexity of reality to a unified theoretical framework. In matter of time, therefore, the world can be investigated according to several interpretations, and it does not seem possible to merge them into a ultimate theory: the recent synthesis of quantum gravity, though it is founded on a model in which the contradictions between quantum mechanics and general relativity are, in some respects, brilliantly overcome, does not solve in fact the problem of time, of which it merely contemplates, at a fundamental level, the not existence, in the illusion that such reductionist vision can explain its ultimate nature. We believe that the problem does not consist in the possibility of explaining the emergence of irreversibility from a reversible timelessness, but in recognizing as autonomous and not always in communication between them the different levels of knowledge of physical phenomena, and consequently the theoretical models used to describe them. What can be deduced, for example, about the reality of relativistic time, since the theory of general relativity provides different decay times of radioactive samples or different aging of biological organisms along different world lines, overlapping two concepts that here we have shown to be of different nature, the reversible relativistic time and the irreversible thermodynamic time? The answer is implicitly contained in the above observations, since currently in physics different levels of investigation around the problem of time are blended, ignoring or forgetting that every field of research produces the conceptual and operational instruments that allow to verify the assumptions on which every particular theory has been founded. If, therefore, relativity requires to be tested through atomic or light clocks, that allow to probe the structure of spacetime on which its conceptual building has been erected, the devices of thermodynamic nature, as radioactive clocks or, more in general, Rovelli's thermal clocks (inside which irreversible phenomena of statistical nature are produced that cannot be explained by the relativistic model), probe that the emerged reality of time is of evolutionary nature, whereby the measure of a duration is linked to an irreversible change that occurs inside the instrument. \\Irreversibility is the peculiar characteristic of physical reality as it appears at the macroscopic scale, where our senses perceive and instruments provide measurements\footnote {To paraphrase the title of a recent Rovelli's essay \cite {rov3}, we can say that the irreversible reality of time appears in thermodynamic form. It must be noted that one of the objectives of theoretical and experimental research should be the description and understanding of reality as it appears, recognizing, in every abstract theorisation, a potential reductionist risk, wherever we want to explain the irreversibility of physical phenomena through the reversible phenomena that seem to constitute their elementary structure.}. Thermodynamic time therefore measures the duration of phenomena that happen at the macroscopic scale, ignoring the microstates, despite these latter, in a more metaphysical than physical sense, can be considered the ultimate and invisible substrate upon which the physical reality is founded. \\Every theoretical idealization, from Newton to Einstein as far as to quantum mechanics and quantum gravity, albeit using different mathematical models as instruments to interpret the complexity of phenomena, is founded on a clear indisputable preconception about a reality of time external to bodies and clocks. 
%%%%%%%%%%%%%%%%%%%%%%%%%%%%%%%%%%%%%%%%%%%%%%%%%%%%%%%%%%%%%%%%%%%%%%%%%%%%%%%%%%%%%%%%%%%%%%%%%%%
\section{Summary and concluding remarks}
The different concepts of time arising from the recent physical theories engender problematic nodes, since, though dialectically, all develop from Newton's theory, in which time is axiomatically conceived as a metaphysical entity, namely a mathematical and therefore not real duration. Relativistic time does not have an absolute nature and its operational definition implies different measures that the different observers provide in relation to their states of motion: in the most complete formalization of the theory, general relativity, the durations are written in the world lines and it is expected, in accordance with the clock hypothesis, that clocks behave as ideal instruments able to measure them. A detailed analysis of real clocks, whose proper period can depend or less on the gravitational or the pseudo gravitational potential, means that ideal clocks do not exists or, however, that real clocks belong to sets irreducible to each other: the fact that only atomic clocks, and a few other equivalent to them, behave in accordance with theory, makes the relativistic theory of time implicitly contradictory. \\Problems and doubts about the matching between the relativistic predictions and the experimental measurements provided by real clocks have been remarked by some authors, in particular by Brown, according to which the behaviour of clocks in general relativity correlates with some aspects of spacetime, so it is not correct to consider them as instruments that measure a physical quantity existing in their inside. \\Radioactive clocks, that measure durations through the quantification of the mass of decayed substance in correspondence to a given phenomenon between two extreme events that mark its beginning and its end (for example, the birth and the decay of an unstable particle), with good probability do not operate in agreement with relativistic laws, and the multiple experimental tests about the different mean lifetime of a radioactive sample have only verified the relationship between the measurements performed by two observers in relative motion, therefore not allowing to conclude anything about the different internal evolution of the measuring devices. Experiments for proving if a radioactive sample, in correspondence to two nonequivalent world lines between two fixed extreme events, registers different amounts of decayed substance, have not yet been performed: only an experimental test that compares the behaviour of radioactive clocks can say something definitive about the nature and the reality of radioactive time\footnote{As claimed by Basri \cite {bas}, clocks whose operation is linked to different forms of interaction do not necessarily behave in an equivalent manner: the equivalence, in the current state of experimental findings, is practically gained for clocks whose operation is related to strong nuclear or electromagnetic interactions, but it is problematic about the behaviour of radioactive (weak nuclear force) or gravitational clocks. Possible abnormalities of behaviour, explicitly recognized also by Finzi \cite {fin}, have never been subjected to a careful investigation in relation to a critical analysis or an operational redefinition of physical time.}. \\The critical analysis allows to deduce that, while Newtonian time is a set of instants of universe, time of special relativity is linked to the procedure of synchronization of clocks, obtained by each observer through the exchange of light signals with other ideal and identical clocks, assuming the isotropy and the invariance of the speed of light. After having filled the space with synchronized clocks, an observer measures the duration of a phenomenon as difference between the instants that correspond to the extreme events. The step towards the fusion between space and time, initially formalized by Minkowski, was completed, in view of a more rigorous operational definition of relativistic time, by Einstein's theory of general relativity, in which the spacetime has an objective physical reality, whose metric is determined by the density of matter-energy. The overcoming of Newtonian metaphysics of absolute space and time has therefore engendered, through Einstein's theory, the metaphysical immanentism of the absolute spacetime, in which the instants are not scanned by a single universal clock that measures the duration of each phenomenon and the life of every body, but are scanned by clocks that record proper durations as lengths of world lines. While in special relativity any clock, belonging to the set that ideally covers the continuum of space, does not quantify a real internal becoming, we have shown, as concerns the conceptual framework of general relativity, that some real clocks do not behave as relativistic clocks, as they record internal evolutions irreducibly different. \\
According to the thermal time hypothesis, the problem can be translated in terms of the probable disagreement between the experimental measurements obtained by relativistic clocks and those provided by thermal clocks. It may be objected, according to Barbour and Rovelli, that a description of fundamental (quantum or
relativistic) physics can be given ignoring time, but the underlying problem, at the conceptual level, remains open and risks to generate other contradictions: what does it mean to predict different lifetimes of a radioactive sample or of a biological organism varying the world line, if the relativistic time is conceived, as Rovelli properly does, as a quantity related to the gravitational or pseudo gravitational potential, since a radioactive or a biological clock records an internal evolution whose rate is very probably independent of this potential? The doubt, expressed by Einstein already in the framework of special relativity, about the consistent response of the measuring instruments, was not solved by the theory of spacetime, opening the way to cosmological theories in which the operational concept of time shows to be intrinsically problematic, and to many twentieth-century mathematical representations of phenomenal reality, in which multidimensional abstract spaces proliferate, turning mind away from nature or confusing nature with mind, as if in the world of phenomena we could meet the abstract realities that we have conceived in the Platonic world of ideas. \\
From the analysis and the considerations developed in this paper we infer the necessity to put a new border line between observational reality and mathematical hypothesis, then between physics and metaphysics, starting from the inconsistencies that can be detectable inside the theories on the basis of objective experimental results. It can be argued that any physical theory is metaphysically built, since, being an intellectual processing that pretends to represent and explain the empirical reality in the light of a logical synthesis, it needs to be founded on axiomatic assumptions. The objection is correct, but it does not prevent us to observe, in matter of time, that a consistent theory of durations must be based on postulates that contain necessary references to measurements obtained by real clocks. Physical and philosophical interpretations of the complexity of phenomena conceived by physicists and epistemologists, from Einstein to Malament and Maudlin, until Rovelli and Barbour, cannot avoid that physics has to take into account the variety of change that microscopic or macroscopic reality shows to the eyes of the experience. \\
The conclusions, whose importance should be stressed, are two. \\
Firstly we have found that only a small number of real clocks operate in agreement with relativistic predictions in matter of measure of durations, from which it can be deduced that the relativistic theory of time does not say anything definitive about the mean lifetime of radioactive samples or, in the likely hypothesis that also biological clocks do not behave as atomic clocks, about the different rate of aging of living organisms, merely providing measurements related to the gravitational or pseudo gravitational potential in which the instrument has been during the course of a phenomenon. \\
Secondly, it should be noted that the theories, albeit referring to the same material of investigation, often are not extensions of each other, but rather representations of reality between them largely irreconcilable. The clear and indisputable evidence, in addition to the limited correspondence between measures obtained by real clocks and relativistic predictions, is given by the substantial difference between the concept of evolutionary internal time and that of proper time which, in hindsight, is already written in the world line and thus does not contain any information about the evolution rate of bodies and clocks that describe it. \\
How is changed, therefore, the representation of reality in the direction of the current theoretical syntheses? Newtonian time leaves intact the possibility of becoming: its absolute being means possibility of open evolution, not determined by the structure of universe. The Einsteinian revolution consists in the relativisation of the concepts of space and time, at the price of creating a model of absolute spacetime, so of a new absolute, albeit of different nature, to be placed at the foundation of real phenomena. Relativistic time is a constituent part of spacetime, whose structure is determined by the density of matter-energy: as such it measures a closed becoming, without actual evolution. Thermodynamics is the only theory in which the evolutionary nature of time is fundamental, since it postulates the existence of an internal time, generated inside real clocks that measure durations through the irreversible phenomena that occur inside them. An accurate analysis of the behaviour of real clocks will lead to a refoundation of the physical thought about time, from which it will probably emerge that this complex concept cannot be explained in the light of a unitary theoretical structure.

\vskip 0.5cm
\textbf{Acknowledgements} Thanks to Carlo Rovelli and Julian Barbour for having discussed some aspects of this theoretical analysis of the problem of time. A special thanks to Silvio Bergia (University of Bologna) for his precious and concrete attention to the development of these ideas.
\vskip 1cm
% ---- Bibliography ----
%

%

\begin{thebibliography}{5}

\bibitem{ein1}
Einstein, A.: Zur Elektrodynamik bewegter Körper, \textit {Annalen der Physik} \textbf {17} (1905)
\bibitem{cat}
Cattaneo, C.: Sui postulati comuni alla cinematica classica e alla cinematica relativistica, \textit {Rend. Acc. Lin.} \textbf {24} (1958) 
\bibitem{pal}
Pal, P.B.: Nothing but relativity, \textit {Eur. J. Phys.} \textbf {24} (2003)
\bibitem{rei}
Reichenbach, H.: Axiomatik der Relativistischen Raum-Zeit-Lehre, Friedrich Vieweg and Son Braunschweig (1924)
\bibitem{bas}
Basri, S.A.: Operational foundation of Einstein’s General theory of relativity, \textit {Review Of  Modern Physics} \textbf {37} (1965)
\bibitem{haf}
Hafele, J.C.,  Keating, R.E.: Around-the-World Atomic Clocks: Predicted Relativistic Time Gains, \textit {Science} \textbf {177} (1972)
\bibitem{all}
Alley, C.O.: Relativity and clocks, Proc. XXXIII Ann. Symp. on Frequency Control 4-39; http://dx.doi.org/10.1109/FREQ.1979.200296 (1979)
\bibitem{ash}
Ashby, N.:  Relativistic Effects in the Global Positioning System, \textit {Living Rev. Relativity} \textbf {6} (2003)
\bibitem{rov1} 
Rovelli, C.: Forget time, arXiv:0903.3832v3 (2008)
\bibitem{bel}
Bell, J.S.: How to teach special relativity, Progress in Scientific Culture (1976), reprinted in J.S. Bell, Speakable and Unspeakable in Quantum Mechanics, Cambridge: Cambridge University Press, 67-80 (2008)
\bibitem{pau}
Pauli, W.: Relativitatstheorie, Encyklopaedie der matematischen Wissenschaften, mit Einschluss ihrer Anwendungen vol 5 - Physik (1921)
\bibitem{bro}
Brown, H.: The behaviour of rods and clocks in general relativity, and the meaning of the metric field, arXiv:0911.4440v1 [gr-qc] (2009)
\bibitem{kno}
Knox, E.: Flavour-Oscillation Clocks and the Geometricity of General Relativity, arXiv:0809.0274v1 (2008) 
\bibitem{ahl}
Ahluwalia, D.V.: On a New Non-Geometric Element in Gravity”,  arXiv:gr-qc/9705050v2 (1998)
\bibitem{mal}
Malament, D.: Classical relativity theory, Elsevier (2006)
\bibitem{mau}
Maudlin, T.: Philosophy of physics: space and time, Princeton University Press (2012)
\bibitem{kos}
Kostro, L.:  What is this: a clock in relativistic theory? in Recent advances in relativity theory (2000)
\bibitem{bri}
Briatore, L.,  Leschiutta, S.: Evidence for the earth gravitational shift by direct atomic-time-scale comparison, \textit {Nuovo Cimento B} \textbf {37} (1977)
\bibitem{bor1}
Borghi, C.: Clock effect and operational definitions of time, \textit{Annales Fond. Broglie} \textbf{37}, 227 (2012) 
\bibitem{bor2}
Borghi, C.: Hypothesis about the nature of time and rate of clocks, \textit{Annales Fond. Broglie} \textbf{38}, 167 (2013)
\bibitem{bor3}
Borghi, C.: Are mechanical clocks relativistic clocks?, \textit{Annales Fond. Broglie} \textbf{39}, 95 (2014)
\bibitem{bor4}
Borghi, C.: Physical time and thermal clocks, \textit{Found. Phys.} \textbf{46} (2016)
\bibitem{bai}
Bailey, J., \textit {et al}: Measurements of Relativistic Time Dilatation for Positive and Negative Muons in a Circular Orbit, \textit {Nature} \textbf {268} (1977)
\bibitem{eis}
Eisele, A.: On the behaviour of an accelerated clock, \textit {Helvetica Physica Acta} \textbf {60} (1987)
\bibitem{wil}
Will, C.M.: The Confrontation between General Relativity and Experiment, \textit {Living Rev. Relativity}, \textbf {9} (2006)
\bibitem{ein2}
Einstein, A.: Autobiographical Notes, Chicago: Open Court Publishing Company (1949)
\bibitem{bar}
Barbour, J.: The nature of time,  arxiv.org/pdf/0903.3489 (2008)
\bibitem{rov2}
Rovelli, C.: Quantum Gravity, Cambridge University Press (2004)
\bibitem{rug}
Rugh, S.E., Zinkernagel, H.: On the physical basis of cosmic time,  arXiv:0805.1947v1 [gr-qc] (2008)
\bibitem{rov3}
Rovelli, C.: Reality is not as it appears, Raffaello Cortina Editore (2014)
\bibitem{fin}
Finzi, A.:  Dimensionless quantities, spacelike intervals and proper time in general relativity, \textit {Nuovo Cimento} \textbf {20} (1961)


\end{thebibliography}
\end{document}